\documentclass[
 a4paper,
 10pt,
 superscriptaddress,
 amsmath,amssymb,
 twocolumn,
 aps,
 pra,
 longbibliography,
 floatfix,
 notitlepage,
 showpacs,
 noeprint,
]{revtex4-2}

\usepackage{braket}
\usepackage[colorlinks=true,citecolor=blue]{hyperref}
\usepackage[normalem]{ulem}
\usepackage[english]{babel}
\usepackage{xcolor,graphicx}

\DeclareMathOperator{\arctanh}{arctanh}
\newcommand{\fpm}[1]%
{\begingroup{\color{red} FPM: \textsl{#1}}\endgroup}

\begin{document}

\title{Quantum limits on squeezing}

\author{Francesco Massel}
 \affiliation{Department of Science and Industry Systems, University of
   South-Eastern Norway, P.O. Box 235 Kongsberg, Norway}
 \affiliation{Department of Physics, University of Oslo, Oslo N-0316, Norway}

 \begin{abstract}
   In our work, we show how, for a network of bosonic modes, canonical
   commutation relations constrain the coefficients relating input and internal
   modes. Based on these constraints, we derive a lower bound on the total
   steady-state squeezing achievable in reservoir-engineered (dissipative)
   squeezing schemes, quantified by the sum of mode-optimal quadrature variances
   each normalized to its corresponding input variance. The bound on this
   normalized sum follows solely from
   canonical commutation relations and stability, and is saturated in the
   strong-coupling limit at unity. Furthermore, we show that adding independent
   parametric driving terms for each mode changes the quantum noise--gain balance
   and yields, for the same normalized mode-optimal sum, a distinct optimum
   bound, approaching 1/2. In addition, we show
   how these constraints allow us to reformulate the Duan inseparability
   criterion for a three-mode bosonic system in terms of a single
   parameter-dependent figure of merit. Our results apply directly to current
   electromechanical and nanomechanical experiments and indicate that the
   two-mode bounds can be experimentally approached even at room temperature.
 \end{abstract}
\maketitle

Constraints imposed by the laws of quantum mechanics can shape the macroscopic behavior of physical systems, from the Pauli exclusion principle underlying the chemical properties of matter, to the fundamental limits on measurement and control~\cite{penrose_gravitys_1996,zurek_decoherence_2003}. In linear bosonic devices, noncommutativity enforces unavoidable tradeoffs, exemplified by the quantum limit on added noise in amplification~\cite{haus_quantum_1962,caves_quantum_1982,caves_quantum_2012-1} and by the 3~dB limit for stable intracavity parametric squeezing~\cite{collett_squeezing_1984,collett_squeezing_1985,clerk_introduction_2010}.

Here we show that, for a stable physically realizable Markovian network of $N$ bosonic modes, canonical commutation relations constrain the transfer coefficients between input and internal modes through simple commutator-budget sum rules. For passive (number-conserving) linear networks---including reservoir-engineered squeezing schemes that become passive in an appropriate Bogoliubov frame---these commutator budgets also act as noise budgets, directly constraining steady-state quadrature variances and thus the squeezing achievable by engineered dissipation.

While our framework applies to generic linear bosonic networks, we focus on examples motivated by cavity optomechanics and electromechanics\cite{aspelmeyer_cavity_2014}, where controllable radiation-pressure interactions enable reservoir engineering of dissipation and interactions~\cite{poyatos_quantum_1996, metelmann_quantum-limited_2014,metelmann_nonreciprocal_2015,pokharel_coupling_2022}. On the theory side, engineered dissipation provides a versatile route to prepare nonclassical steady states, including ground-state cooling~\cite{genes_ground-state_2008}, steady-state mechanical squeezing~\cite{kronwald_arbitrarily_2013}, and stationary optomechanical and mechanical--mechanical entanglement~\cite{mari_gently_2009,wang_reservoir-engineered_2013,tan_achieving_2013,woolley_two-mode_2014}. Experimentally, reservoir-engineering techniques have enabled ground-state cooling of nanomechanical and micromechanical resonators~\cite{teufel_sideband_2011-1,chan_laser_2011}, observation of steady-state mechanical squeezing~\cite{wollman_quantum_2015,lecocq_quantum_2015,pirkkalainen_squeezing_2015}, and stabilized entanglement involving mechanical degrees of freedom~\cite{palomaki_entangling_2013,ockeloen-korppi_stabilized_2018}.

These constraints yield a lower bound on the total steady-state squeezing achievable by two-mode reservoir engineering~\cite{kronwald_arbitrarily_2013,woolley_two-mode_2014}: the sum of the two mode-optimal quadrature variances (each normalized to its corresponding input variance) cannot be reduced below $1$, and approaches this limit in the strong-coupling regime. We further show that adding local degenerate parametric driving modifies the noise--gain balance and leads to an optimized bound approaching $1/2$ within the stability region.

Finally, we apply the same commutator-budget framework to the three-mode reservoir-engineering scheme of Refs.~\cite{woolley_two-mode_2014,ockeloen-korppi_stabilized_2018} and recast the Duan inseparability criterion~\cite{duan_inseparability_2000} in terms of a single commutator-transfer parameter, enabling a direct assessment of entanglement performance from system parameters.

Our starting point is the description of an $N$-mode bosonic system coupled to an external environment, which we will describe with standard input--output theory~\cite{gardiner_input_1985}. We suppose that each mode is independently coupled to its bath, which determines noise and dissipation properties of the corresponding bath, and consider a  time evolution governed by a linear time-invariant (LTI) system of equations which, in the Bogoliubov (doubled)
space, can be written as $ \dot{\xi} = \bar{A} \xi + D \xi_\mathrm{in}$~\cite{james_hinfty_2008,nurdin_linear_2017}, where
$\xi=\left[a_1,a_2\dots a_N;a^{\dag}_1,a^{\dag}_2\dots a^{\dag}_N \right]$
($\xi_\mathrm{in}=\left[a_{1,\mathrm{in}},a_{2,\mathrm{in}}\dots a_{N,\mathrm{in}};a^{\dag}_{1,\mathrm{in}},a^{\dag}_{2,\mathrm{in}}\dots a^{\dag}_{N,\mathrm{in}}\right]$),
$\bar{A}$ and $D$ are the drift and input matrices, respectively. We assume all dissipation is accounted for by these explicit Markov bath couplings (i.e., no additional unobserved/internal loss channels), so that input channels are in one-to-one correspondence with damped modes.

Under these assumptions, the relations derived below follow solely from stability ($\operatorname{Re}\lambda(\bar A)<0$) and the preservation of canonical commutation relations (CCRs), which, following~\cite{james_hinfty_2008}, we term  physical realizability (PR)~\cite{noauthor_supplemental_2026}.

For a PR system, it is
possible to define a $2\mathrm{N} \times 2\mathrm{N}$ matrix
\begin{equation}
  \label{eq:1}
  W_i = \int_0^{\infty} e^{\bar{A}t} (D_{i} \Sigma D_{i}^\dagger) e^{\bar{A}^\dagger t} dt
\end{equation}
where $\Sigma= [I_N,\, 0; 0,\, \text{-}I_N]$, and $D_i$ denotes the columns of $D$ associated with input channel $i$. ${(W_i)}_{jk}$ quantifies the
contribution of input channel $i$ to the commutator budget of the commutator
$\left[\xi_j, \xi^{\dag}_k\right]$ for modes $j$ and $k$ in Bogoliubov space.
From Eq.~\eqref{eq:1}, it is possible to show that, for stable $\bar{A}$, $W_i$ is the unique solution to
the Lyapunov equation $\bar A W_i + W_i \bar A^\dagger + D_i \Sigma D_i^\dagger=0$. We can restrict $W_i$  to the annihilation sector defining the
($\mathrm{N} \times \mathrm{N}$) commutator channel associated with input $i$ as the projection $K_i=P_a W_i P_a^\dag$, where $P_a = [ I_N, 0 ]$ (size
$\mathrm{N} \times 2\mathrm{N}$).

PR for internal modes $\left[a_j,a_k^\dagger\right] = \delta_{jk}$ immediately
implies that $\sum_i ~{(K_i)}_{jk}=\delta_{jk}$. In addition, for passive
systems, it is possible to show that $K_i$ is a positive matrix. If we further
consider the usual situation in which dissipation is diagonal (i.e.
the annihilation sector $A+A^\dag=-\mathrm{diag}(\gamma_1,\dots,\gamma_N)$), and
internal dissipation is absent, we have that $\sum_j \gamma_j (K_i)_{jj} = \gamma_i$. Upon
further restricting to real couplings (i.e. to reciprocal
systems~\cite{metelmann_nonreciprocal_2015}) we have
$\gamma_j (K_i)_{jj}=\gamma_i (K_j)_{ii}$ which can be considered as a manifestation of
Onsager reciprocity~\cite{onsager_reciprocal_1931,noauthor_supplemental_2026}.
For the $N=2$ case, any (generally complex) Hermitian matrix can be transformed
to a real one, making the reciprocity condition redundant. This can be confirmed
by showing how, for $N=2$, reciprocity can be directly deduced from
$\sum_i (K_i)_{jk}=\delta_{jk}$ and $\sum_j \gamma_j (K_i)_{jj}=\gamma_i$.

If the LTI system is solved explicitly we can write the lowering operators
associated with the internal modes as
\begin{align}
  \label{eq:2} a_i = \sum_j \int \frac{d \omega}{2\pi} \left[M_{i,j}(\omega) a_{\mathrm{in},j} + L_{i,j}(\omega) a^\dagger_{\mathrm{in},j}\right],
\end{align}
from which we can deduce
$(K_j)_{ik} = \int \frac{d \omega}{2\pi} \left[{M^*}_{i,j}(\omega) {M}_{k,j}(\omega)-{L^*}_{i,j}(\omega) {L}_{k,j}(\omega)\right]$.
For passive systems ($L_{i,j}=0$), it immediately follows that, for uncorrelated
inputs, the steady-state equal-time variance
$\Delta X_i^2 = \braket{X_i^2}- \braket{X_i}^2$, where
$X_i(\theta)=\left(a^\dag e^{i\theta}+a e^{-i\theta}\right)/\sqrt{2}$ ($X_i \equiv X_i(\theta)$, $P_i\equiv X_i(\pi/2)$ ), can be expressed in terms of the inputs as $\Delta X_i^2 = \sum_j (K_j)_{ii} \Delta X_{j,\mathrm{in}}^2$. The passivity of the system therefore allows us to establish a crucial link between CCR constraints and power/noise gain constraints in a linear bosonic system.

The full matrices \(K_i\) encode diagonal and off-diagonal commutators \([a_j,a_k^\dagger]\) and are required when treating inter-mode correlations. In the following we focus on their diagonal terms, which we denote \(I_{ij}=(K_j)_{ii}\). \(I_{ij}\) determines the contribution of input \(j\) to the equal-time commutator (and, for passive systems with uncorrelated inputs, to the local quadrature variance of mode \(i\)).

\paragraph*{Dissipative squeezing} The first case we consider is a coupled
two-mode system in the presence of diagonal dissipation. We will show how the
relations established above allow us to determine nontrivial limits on the
quadrature variances in a dissipative squeezing scenario.

The Hamiltonian we are considering describes a standard reservoir engineering
setup in which two bosonic modes are coupled through the following (interaction)
Hamiltonian
\begin{align}
  \label{eq:3} H = G_+ a_1^\dagger a^\dagger_2 + G_- a_1^\dagger a_2 + \mathrm{h.c.}.
\end{align}
Eq.~\eqref{eq:3} represents, for instance, the prototypical optomechanical setup
leading to squeezing of mechanical
motion~\cite{wollman_quantum_2015,pirkkalainen_squeezing_2015,lecocq_quantum_2015}.
The Hamiltonian~\eqref{eq:3} can be transformed as
$ H = \mathcal{G} a_1^\dagger \alpha + \mathrm{h.c.}$ where the Bogoliubov mode
$\alpha$ is defined as $\alpha= \cosh[\xi] a_2 + \sinh[\xi] a_2^\dagger$,
$\mathcal{G}=\sqrt{G_-^2-G_+^2}$ and $\xi =\arctanh\left[G_+/G_-\right]$, with
$|G_+|<|G_-|$. Furthermore, if we assume the dissipative properties of such
system to be described by the conventional coupling of each mode to a Markovian
(thermal) bath, we realize that, in the Bogoliubov basis $(a_1,\alpha)$ the
interaction is beam-splitter type, and hence the system fulfills the conditions
stated above (passivity, diagonal dissipation, no hidden dissipation channels),
allowing us to write $I_{ii}+I_{ij}=1$ and
$\gamma_i I_{ii}+ \gamma_j I_{ji}=\gamma_i$ with $i,j \in {1,b}$ (where $b$ is
the index associated with the Bogoliubov mode). From these relations we can
define $I_x = I_{1b}/\gamma_b = I_{b1}/\gamma_1,\qquad \gamma_b=\gamma_2$, since
the losses associated with the Bogoliubov mode $b$ are equal to the losses in
the original basis ($\gamma_2$).

For a two-mode system in the presence of coherent coupling one can show that $I_x \leq 1/(\gamma_1+\gamma_2)$\cite{noauthor_supplemental_2026} and therefore
$I_{ii} \geq \gamma_i/(\gamma_i+\gamma_j)$ and $I_{11}+I_{22} \geq 1$.
Remembering that $\Delta X_i^2 = \sum_{j} I_{ij} \Delta X^2_{j,in}$, we can conclude that
$\Delta X_1^2(\theta)/ \Delta X_{1,in}^2 + \Delta X_{\alpha}^2(\theta)/\Delta X_{\alpha,in}^2(\theta) \geq 1$. In the Bogoliubov frame, the effective input $\alpha_{\mathrm{in}}$ is a Bogoliubov transform of a thermal bath and therefore has $\braket{\alpha_{\mathrm{in}}\alpha_{\mathrm{in}}}\ne 0$. As a result, the quadrature variances of both modes become $\theta$ dependent.
This phase dependence affects $\Delta X_1$ as well, since mode 1 depends on both
inputs. Since the inequality is fulfilled for any phase $\theta$, it holds in
particular for minimal-variance quadrature. Transforming
$\alpha$ back to the original $a_2$ mode, allows us to conclude that
\begin{equation}
  \label{eq:4}
  \left. \Delta X_1^2/ \Delta X_{1,in}^2\right|_{min} + \left. \Delta X_{2}^2/\Delta X_{2,in}^2\right|_{min} \geq 1,
\end{equation}
indicating that the overall ``squeezing power'' (i.e. sum of the ratios between
system and input variances for both modes) is bounded from below by 1. It
is not possible to squeeze both modes of a system described by Eq.~\eqref{eq:3}
in the presence of Markovian coupling to thermal baths below a threshold
imposed, ultimately, by CCR.
\begin{figure}[tbp]
  \centering
  \includegraphics[width=\columnwidth]{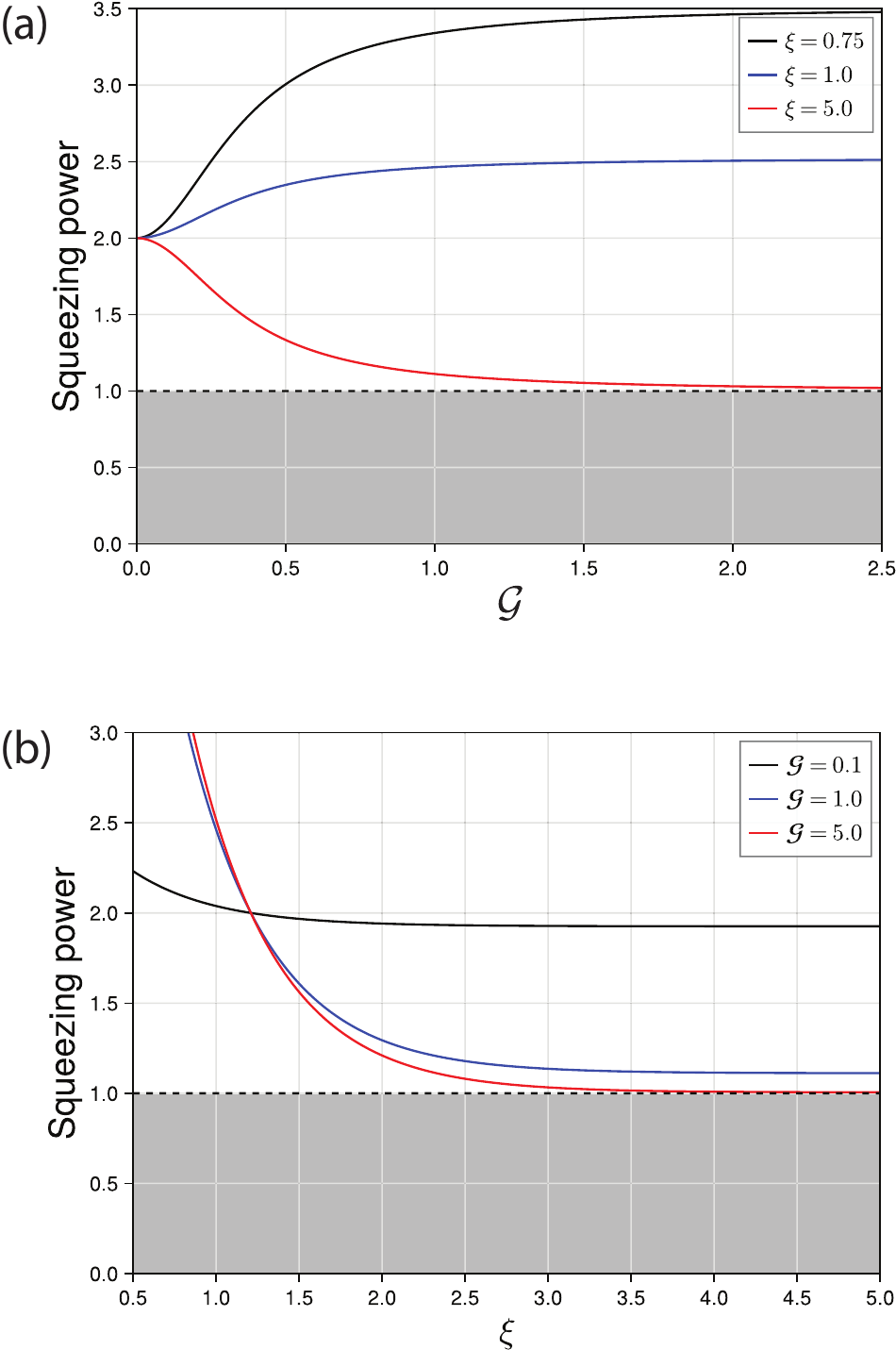}
  \caption{Squeezing power of a two-mode purely dissipative setup. (a) Sum of the two mode-optimal quadrature variances, normalized to their corresponding input variances, as a function of the coherent coupling strength and the Bogoliubov squeezing parameter. (b) Cuts of the same quantity for representative values of the Bogoliubov squeezing parameter, showing the approach to the lower bound in the strong-coupling limit.}
  \label{fig:1}
\end{figure}

\begin{figure}[tbp]
  \centering
  \includegraphics[width=\columnwidth]{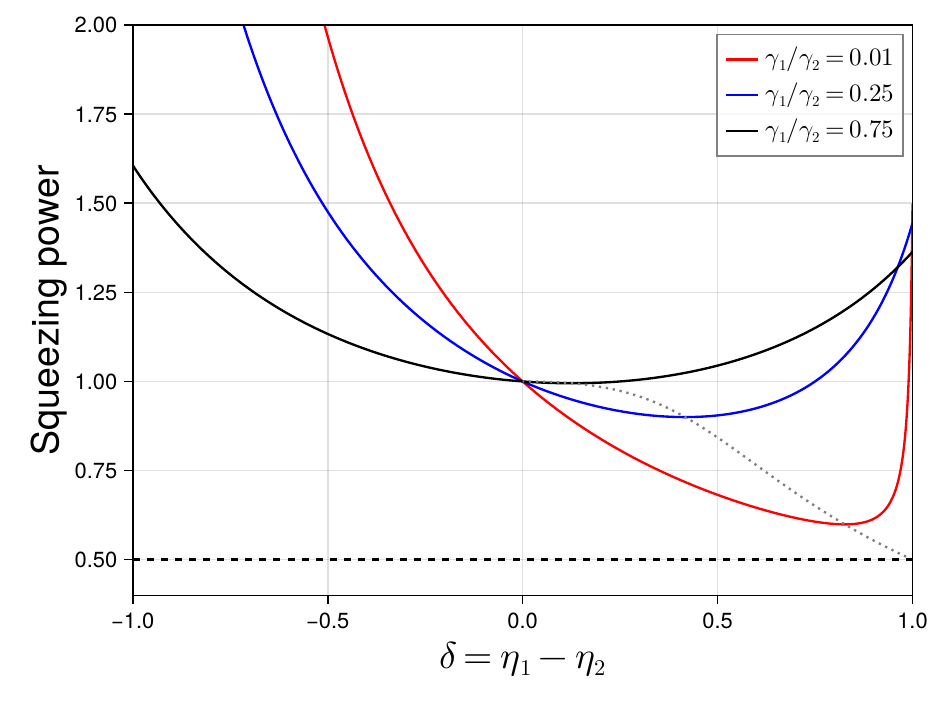}
  \caption{Squeezing power in the presence of degenerate parametric terms. The  dotted line corresponds to the minimum attainable squeezing power as a function of $\delta=\eta_1-\eta_2$ for different values of ratio $\gamma_1/\gamma_2$.}
  \label{fig:2}
\end{figure}

The analysis can be extended to include degenerate
parametric terms
\begin{align}
  \label{eq:5} H = G_+ a_1^\dagger a_2 + G_- a_1 a_2 + \eta_1 a_1^2 + \eta_2 a_2^2 + \mathrm{h.c.}.
\end{align}
From Eq.~\eqref{eq:5} we can deduce the following equations of
motion for the quadratures $X_{1,2} = (a_{1,2}^\dagger + a_{1,2})/\sqrt{2}$,
$Y_{1,2} = i (a_{1,2}^\dagger - a_{1,2})/\sqrt{2}$
\begin{align}
    \label{eq:6} \frac{d}{dt}
    \begin{bmatrix} X_1\\ Y_2
    \end{bmatrix} = \underbrace{\begin{bmatrix} -\gamma_{1\, -}/2 & \ G_\Delta\\ - G_\Sigma & -\gamma_{2+}/2
    \end{bmatrix}}_{A_1}
    \begin{bmatrix} X_1\\ Y_2
    \end{bmatrix} +
    \begin{bmatrix} \sqrt{\gamma_1}\,X_{1,\mathrm{in}}\\ \sqrt{\gamma_2}\,Y_{2,\mathrm{in}}
    \end{bmatrix}
  \end{align} and
  \begin{align}
    \label{eq:7} \frac{d}{dt}
    \begin{bmatrix} X_2 \\ Y_1
    \end{bmatrix} = \underbrace{ \begin{bmatrix} -\gamma_{2\, -}/2 & \ G_\Delta\\ - G_\Sigma & -\gamma_{1+}/2
    \end{bmatrix}}_{A_2}
    \begin{bmatrix} X_2 \\ Y_1
    \end{bmatrix} +
    \begin{bmatrix} \sqrt{\gamma_2}\,X_{2,\mathrm{in}}\\ \sqrt{\gamma_1}\,Y_{1,\mathrm{in}}
    \end{bmatrix}
  \end{align} where $\gamma_{1\,\pm}= \gamma_1 \pm \eta_1 $,
$\gamma_{2\,\pm}= \gamma_2 \pm \eta_2$, $G_\Sigma = G_- + G_+$ and
$G_\Delta =G_-- G_+$. Since Eqs.~(\ref{eq:6},\ref{eq:7}) are block-diagonalized
into blocks that involve commuting operators only (quantum mechanics-free subspaces \cite{tsang_coherent_2010,tsang_evading_2012}), we can assume that, when considered in isolation, each quadrature dynamics is generated by a
free Hamiltonian in the presence of coupling to an engineered bath. For example,
the dynamics of the $X_1$ quadrature, for $G_\pm =0$, can be written as
\begin{align}
  \label{eq:8} \dot{X}_1 =-\frac{\gamma_{1-}}{2} X_1 + \sqrt{\gamma_{1-}} \tilde{X}_{1,in}
\end{align} where we have defined
$\tilde{X}_{1,in} = \sqrt{\gamma_1 /\gamma_{1-}}X_{1,in}$, and analogously for the other quadratures. Relying on the fact that the bounds established for $I_{ij}$ depend
on the nature of the coupling coefficients and not on the state of the baths,
Eq.~\eqref{eq:8} and its analogues for the other quadratures allow us to recast
the ``squeezing power'' condition for a system in the presence of degenerate
parametric terms in the same frame of the problem where only dissipative
squeezing is present, in particular
$\Delta X_{1,2}^2 \geq \gamma_{1,2-} /( \gamma_{1,2-} + \gamma_{2,1+}) \Delta \tilde{X}_{1,2,in}^2 $
and
$\Delta Y_{1,2}^2 \geq \gamma_{1,2-} /( \gamma_{1,2-} + \gamma_{2,1+}) \Delta \tilde{Y}_{1,2,in}^2 $,
which can be expressed in terms of the original input variances as
$\Delta X_{1,2}^2 \geq \gamma_{1,2} /( \gamma_{1,2-} + \gamma_{2,1+}) \Delta X_{1,2,in}^2 $ and
$\Delta Y_{1,2}^2 \geq \gamma_{1,2-} /( \gamma_{1,2-} + \gamma_{2,1+}) \Delta Y_{1,2,in}^2 $.
In this case, the boundary for the overall squeezing power becomes
\begin{align}
  \label{eq:9} \frac{\Delta X_{1}^2}{\Delta X_{1,in}^2} + \frac{\Delta X_{2}^2}{\Delta X_{2,in}^2} \geq \frac{\left(\gamma_1 + \gamma_2 \right)^2 - (\eta_1-\eta_2)(\gamma_1-\gamma_2)}{\left(\gamma_1 + \gamma_2 \right)^2- (\eta_1-\eta_2)^2}
\end{align}
which, upon minimization yields a minimum value (within the stability
region of the system) $1/2 + \sqrt{\gamma_1 \gamma_2}/(\gamma_1+\gamma_2)$ for
$\eta_1- \eta_2 =\left(\gamma_1 +\gamma_2\right)(\sqrt{\gamma_1}-\sqrt{\gamma_2})/(\sqrt{\gamma_1}+\sqrt{\gamma_2})$,
which approaches $1/2$ from above as $\gamma_1/\gamma_2 \to 0$ or $\gamma_2/\gamma_1 \to 0$ (see Fig.~\ref{fig:2}). In this limit, one mode-optimal normalized quadrature variance can approach zero, while the other approaches $1/2$.

While this discussion does not affect the usual dissipative optomechanical
setup, since the optical linewidth is much larger than the mechanical one,
resulting in an extreme imbalance of the squeezing power
$\Delta X_{opt}^2 / \Delta X_{opt,in}^2 \simeq 1$, $\Delta X_{mech}^2 / \Delta X_{mech,in}^2 \gtrsim 0$,
our analysis is relevant for purely mechanical realizations of dissipative
squeezing~\cite{xinexperiment}.

\paragraph{3-mode system.} In the following, we will discuss a three-mode optomechanical setup,  corresponding to the situation of Ref.~\cite{woolley_two-mode_2014, ockeloen-korppi_stabilized_2018}, where entanglement between two mechanical  oscillators was obtained by means of dissipative squeezing of collective mechanical modes, mediated by a microwave cavity. Our starting point is the analysis performed in Refs.~\cite{woolley_two-mode_2014, ockeloen-korppi_stabilized_2018} where  entanglement of modes 2 and 3 is verified by ascertaining the violation of the Duan bound~\cite{duan_inseparability_2000}, stating that, for a separable state, the following inequality must be fulfilled $\Delta X_\Sigma^2 + \Delta P_\Delta^2 \geq 1$.  $\Delta X_\Sigma^2$ and $\Delta P_\Delta^2$ are the variances of the
symmetric and antisymmetric quadratures $X_\Sigma= (X_2 + X_3)/\sqrt2$,
$P_\Delta = (P_2 - P_3)/\sqrt2$.

Here we show that the separability criterion of the two mechanical modes can be formulated in terms of one of the transfer integrals $I_{ij}$ introduced above and the squeezing parameter, allowing for a simple direct verification of entanglement from the knowledge of the system parameters.

Considering a somewhat simplified description (e.g. validity of the rotating-wave approximation, frequency of the sideband drives perfectly matched), we can write the Hamiltonian (in the appropriate frame) for the system as
\begin{multline}
  \label{eq:10} H= G_- a_1^\dagger \left( a_2 +a_3 \right) +G_+ a_1^\dagger\left( a_2^\dagger + a_3^\dagger\right) + \mathrm{h.c.} \\ + \Omega \left(a_3^\dagger a_3 - a_2^\dagger a_2\right)
\end{multline}
where 1 is the optical mode, 2 and 3 are the mechanical modes, $G_+$ and $G_-$ are the blue- and red-sideband coherent drives, and $\Omega=(\omega_3-\omega_2)/2$ is half the frequency difference between the mechanical modes, with $\omega_3>\omega_2$.
As for the case of the dissipative squeezing Hamiltonian, it is possible to
perform Bogoliubov transformations, for modes 2 and 3. This allows us to turn
the Hamiltonian in Eq.~\eqref{eq:10} into
\begin{equation}
  \label{eq:11}
  H= \sqrt{2}\,\mathcal{G} a_1^\dagger \alpha_\Sigma- \Omega \alpha_\Sigma\alpha^\dagger_\Delta + \mathrm{h.c.}
\end{equation}
where $\mathcal{G}=\sqrt{G_-^2-G_+^2}$, $\tanh\xi=G_+/G_-$, and $\alpha_{\Sigma,\Delta} = (\alpha_2 \pm \alpha_3)/\sqrt{2}$,
$\alpha_2 = a_2 \cosh[\xi] + a_3^\dagger \sinh[\xi]$,
$\alpha_3 = a_3 \cosh[\xi] + a_2^\dagger \sinh[\xi]$. The solution of the dynamics generated by
Eq.~\eqref{eq:11}, when recast in the original frame, can be written as
$\Delta X_{i}^2 = \sum_j I_{ij}\Delta X_{j}^2 + I_{i1} \Delta X_1^2 e^{\pm 2\xi}$,
$\Delta P_{i}^2 = \sum_j I_{ij}\Delta P_{j}^2 + I_{i1} \Delta P_1^2 e^{\mp 2\xi}$
($i,j = \Delta, \Sigma$) (see~\cite{noauthor_supplemental_2026}). From the previous relations, for equal thermal populations $n_{\mathrm{m}}$ of modes $\Sigma$ and $\Delta$, the Duan bound for the optimally squeezed quadratures can be reformulated in terms of $I_{ij}$ as
\begin{multline}
  \label{eq:12}
\left(I_{\Delta\Sigma}+I_{\Delta\Delta}+I_{\Sigma\Sigma}+I_{\Sigma\Delta}\right)(n_{\mathrm{m}}+1/2) +\\ \left(I_{\Delta1}+I_{\Sigma1} \right)(n_\mathrm{o}+1/2)e^{-2\xi} \geq 1
\end{multline}
where $n_\mathrm{m}$ and $n_\mathrm{o}$ are the occupation of the mechanical modes 2 and 3 (assumed to be equal) and the occupation of the cavity mode $1$, respectively.
If we assume equal dissipation rates for modes $\Sigma$ and $\Delta$ ($\gamma_m$), the two sum rules $\sum_j I_{ij} =1$ and $\sum_i \gamma_i I_{ij}=\gamma_j$ allow us to write the boundary between entangled and separable regions in the $(n_o,n_m)$ plane as
\begin{equation}
  \label{eq:13}
  n_m = - \frac{\eta_e}{2-\eta_e}e^{-2 \xi} \left[ n_o -\left(e^{2 \xi}-1\right)/2 \right]
\end{equation}
where $\eta_e =\eta_e(\mathcal{G}) = \gamma_1/\gamma_m \left[1-I_{11}(\mathcal{G})\right]$ quantifies the fraction of the optical commutator budget transferred into the mechanical collective subspace (\cite{noauthor_supplemental_2026}). For fixed $\xi$, optimizing over $\mathcal{G}$ enlarges the entangled region, and the intercept of the line with the $n_o$ axis $\bar{n}_o=\left(e^{2 \xi}-1\right)/2$ is entirely determined by the squeezing parameter $\xi$. The (negative) slope is given by  $-\frac{\eta_e}{2-\eta_e}e^{-2 \xi}$ ($0\leq \eta_e \leq 2$), leading to a value of the intercept with the $n_m$ axis $\bar{n}_m=\eta_e\left(1-\exp\left[-2\xi\right]\right)/\left[2\left(2-\eta_e\right)\right]$.
Maximization of $\frac{\eta_e}{2-\eta_e}$, keeping all other experimental parameters fixed, is achieved for an optimal value of $\mathcal{G}_{opt}^2 \simeq \Omega/2  \sqrt{\kappa^2+ 4 \Omega^2}$.

\begin{figure}[tbp]
  \centering
  \includegraphics[width=\columnwidth]{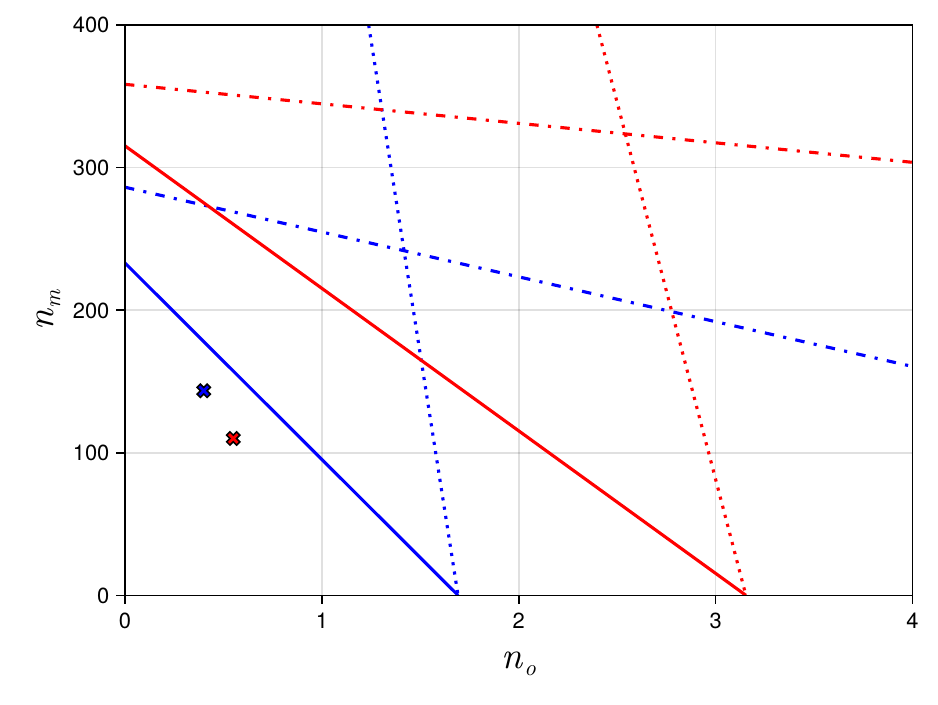}
  \caption{Entanglement limits for the experimental parameters of~\cite{ockeloen-korppi_stabilized_2018}. The blue and red curves show the separability boundaries for datasets A and B of Ref.~\cite{ockeloen-korppi_stabilized_2018}, respectively. Continuous lines depict the boundary for the parameters used in the experiment $\mathcal{G} = \mathcal{G}_{exp}$, dotted lines indicate the boundary for $\mathcal{G} = \mathcal{G}_{opt}$ (with $\xi = \xi_{exp}$), and dash-dotted lines correspond to  $\mathcal{G} = \mathcal{G}_{exp}$ and $\xi = 2\xi_{exp}$. The crosses indicate the actual population of the baths in the experiment.
  }
  \label{fig:3}
\end{figure}

In our work, we established a direct connection between canonical commutation relations and input--output transfer properties for a class of stable, physically realizable linear bosonic systems. For two-mode reservoir-engineered squeezing, these constraints impose a lower bound on the total squeezing power: the sum of the two mode-optimal quadrature variances (each normalized to its corresponding input variance) cannot be reduced below $1$, and approaches this limit in the strong-coupling regime. In the presence of degenerate parametric driving, the optimal bound is modified and can approach $1/2$ within the stability region. Finally, for the three-mode reservoir-engineering scheme of Refs.~\cite{woolley_two-mode_2014,ockeloen-korppi_stabilized_2018}, we recast the Duan inseparability criterion in terms of a single commutator-transfer parameter $\eta_e$, enabling a direct assessment of entanglement performance and optimality from system parameters.

\begin{acknowledgments}
The author thanks Xin Zhou for discussions that motivated this work and concerned its potential experimental implementation. F.M. acknowledges financial support from the Research Council of Norway (Grant No.~333937) through participation in the QuantERA ERA-NET Cofund in Quantum Technologies.
\end{acknowledgments}

\section*{Data Availability}
There are no publicly available research data or software supporting this manuscript. Requests for further information or data should be sent to the author.

\bibliography{SqueezingLimitsv5Zotero}

@article{aspelmeyer_cavity_2014,
  title = {Cavity Optomechanics},
  author = {Aspelmeyer, Markus and Kippenberg, Tobias J and Marquardt, Florian},
  year = 2014,
  month = dec,
  journal = {Reviews of Modern Physics},
  volume = {86},
  number = {4},
  pages = {1391--1452},
  doi = {10.1103/revmodphys.86.1391},
  langid = {english}
}

@article{caves_quantum_1982,
  title = {Quantum Limits on Noise in Linear Amplifiers},
  author = {Caves, Carlton M.},
  year = 1982,
  month = oct,
  journal = {Physical Review D},
  volume = {26},
  number = {8},
  pages = {1817--1839},
  issn = {0556-2821},
  doi = {10.1103/PhysRevD.26.1817},
  urldate = {2025-05-23},
  copyright = {http://link.aps.org/licenses/aps-default-license},
  langid = {english}
}

@article{caves_quantum_2012-1,
  title = {Quantum Limits on Phase-Preserving Linear Amplifiers},
  author = {Caves, Carlton M. and Combes, Joshua and Jiang, Zhang and Pandey, Shashank},
  year = 2012,
  month = dec,
  journal = {Physical Review A},
  volume = {86},
  number = {6},
  pages = {063802},
  publisher = {American Physical Society},
  doi = {10.1103/PhysRevA.86.063802},
  urldate = {2025-05-23}
}

@article{chan_laser_2011,
  title = {Laser Cooling of a Nanomechanical Oscillator into Its Quantum Ground State},
  author = {Chan, Jasper and Alegre, T P Mayer and {Safavi-Naeini}, Amir H and Hill, Jeff T and Krause, Alex and Gr{\"o}blacher, Simon and Aspelmeyer, Markus and Painter, Oskar},
  year = 2011,
  month = oct,
  journal = {Nature},
  volume = {478},
  number = {7367},
  pages = {89--92},
  doi = {10.1038/nature10461},
  langid = {english}
}

@article{clerk_introduction_2010,
  title = {Introduction to Quantum Noise, Measurement, and Amplification},
  author = {Clerk, A. A. and Devoret, M. H. and Girvin, S. M. and Marquardt, Florian and Schoelkopf, R. J.},
  year = 2010,
  journal = {Reviews of Modern Physics},
  volume = {82},
  number = {2},
  pages = {1155--1208},
  issn = {0034-6861},
  doi = {10.1103/revmodphys.82.1155},
  langid = {english}
}

@article{collett_squeezing_1984,
  title = {Squeezing of Intracavity and Traveling-Wave Light Fields Produced in Parametric Amplification},
  author = {Collett, M. J. and Gardiner, C. W.},
  year = 1984,
  month = sep,
  journal = {Physical Review A},
  volume = {30},
  number = {3},
  pages = {1386--1391},
  issn = {0556-2791},
  doi = {10.1103/PhysRevA.30.1386},
  urldate = {2026-04-20},
  copyright = {http://link.aps.org/licenses/aps-default-license},
  langid = {english}
}

@article{collett_squeezing_1985,
  title = {Squeezing Spectra for Nonlinear Optical Systems},
  author = {Collett, M. J. and Walls, D. F.},
  year = 1985,
  month = nov,
  journal = {Physical Review A},
  volume = {32},
  number = {5},
  pages = {2887--2892},
  publisher = {American Physical Society},
  doi = {10.1103/PhysRevA.32.2887},
  urldate = {2026-04-20}
}

@article{duan_inseparability_2000,
  title = {Inseparability {{Criterion}} for {{Continuous Variable Systems}}},
  author = {Duan, Lu-Ming and Giedke, G and Cirac, J I and Zoller, P},
  year = 2000,
  month = mar,
  journal = {Physical Review Letters},
  volume = {84},
  number = {12},
  pages = {2722--2725},
  doi = {10.1103/physrevlett.84.2722},
  langid = {english}
}

@article{gardiner_input_1985,
  title = {Input and Output in Damped Quantum Systems: {{Quantum}} Stochastic Differential Equations and the Master Equation},
  author = {Gardiner, C W and Collett, M J},
  year = 1985,
  month = jun,
  journal = {Physical Review A},
  volume = {31},
  number = {6},
  pages = {3761--3774},
  publisher = {American Physical Society},
  doi = {10.1103/physreva.31.3761},
  langid = {english}
}

@article{genes_ground-state_2008,
  title = {Ground-State Cooling of a Micromechanical Oscillator: {{Comparing}} Cold Damping and Cavity-Assisted Cooling Schemes},
  shorttitle = {Ground-State Cooling of a Micromechanical Oscillator},
  author = {Genes, C. and Vitali, D. and Tombesi, P. and Gigan, S. and Aspelmeyer, M.},
  year = 2008,
  month = mar,
  journal = {Physical Review A},
  volume = {77},
  number = {3},
  pages = {033804},
  issn = {1050-2947, 1094-1622},
  doi = {10.1103/PhysRevA.77.033804},
  urldate = {2025-09-16},
  copyright = {http://link.aps.org/licenses/aps-default-license},
  langid = {english}
}

@article{haus_quantum_1962,
  title = {Quantum Noise in Linear Amplifiers},
  author = {Haus, Hermann A and Mullen, {\relax JA}},
  year = 1962,
  journal = {Physical Review},
  volume = {128},
  number = {5},
  pages = {2407}
}

@article{james_hinfty_2008,
  title = {H{\textsuperscript{\i nfty}} {{Control}} of {{Linear Quantum Stochastic Systems}}},
  author = {James, Matthew R. and Nurdin, Hendra I. and Petersen, Ian R.},
  year = 2008,
  month = sep,
  journal = {IEEE Transactions on Automatic Control},
  volume = {53},
  number = {8},
  pages = {1787--1803},
  issn = {1558-2523},
  doi = {10.1109/TAC.2008.929378},
  urldate = {2026-02-11}
}

@article{kronwald_arbitrarily_2013,
  title = {Arbitrarily Large Steady-State Bosonic Squeezing via Dissipation},
  author = {Kronwald, Andreas and Marquardt, Florian and Clerk, Aashish A.},
  year = 2013,
  month = dec,
  journal = {Physical Review A},
  volume = {88},
  number = {6},
  pages = {063833},
  issn = {1050-2947, 1094-1622},
  doi = {10.1103/PhysRevA.88.063833},
  urldate = {2026-04-20},
  copyright = {http://link.aps.org/licenses/aps-default-license},
  langid = {english}
}

@article{lecocq_quantum_2015,
  title = {Quantum {{Nondemolition Measurement}} of a {{Nonclassical State}} of a {{Massive Object}}},
  author = {Lecocq, F and Clark, J B and Simmonds, R W and Aumentado, J and Teufel, J D},
  year = 2015,
  journal = {Physical Review X},
  volume = {5},
  number = {4},
  pages = {041037}
}

@article{mari_gently_2009,
  title = {Gently {{Modulating Optomechanical Systems}}},
  author = {Mari, A and Eisert, J},
  year = 2009,
  journal = {Physical Review Letters},
  volume = {103},
  number = {21},
  pages = {213603},
  langid = {english}
}

@article{metelmann_nonreciprocal_2015,
  title = {Nonreciprocal {{Photon Transmission}} and {{Amplification}} via {{Reservoir Engineering}}},
  author = {Metelmann, A and Clerk, Aashish A},
  year = 2015,
  month = jun,
  journal = {Physical Review X},
  volume = {5},
  number = {2},
  pages = {021025},
  doi = {10.1103/physrevx.5.021025},
  langid = {english}
}

@article{metelmann_quantum-limited_2014,
  title = {Quantum-{{Limited Amplification}} via {{Reservoir Engineering}}},
  author = {Metelmann, A and Clerk, Aashish A},
  year = 2014,
  month = apr,
  journal = {Physical Review Letters},
  volume = {112},
  number = {13},
  pages = {133904},
  doi = {10.1103/physrevlett.112.133904},
  langid = {english}
}

@misc{noauthor_supplemental_2026,
  title = {Supplemental Material},
  note = {See the ancillary file \texttt{SqueezingLimitsv5Supp.pdf} accompanying this manuscript},
  year = 2026
}

@book{nurdin_linear_2017,
  title = {Linear {{Dynamical Quantum Systems}}},
  author = {Nurdin, Hendra I and Yamamoto, Naoki},
  year = 2017,
  publisher = {Springer International Publishing},
  doi = {10.1007/978-3-319-55201-9},
  isbn = {978-3-319-55199-9}
}

@article{ockeloen-korppi_stabilized_2018,
  title = {Stabilized Entanglement of Massive Mechanical Oscillators},
  author = {{Ockeloen-Korppi}, C F and Damsk{\"a}gg, E and Pirkkalainen, J M and Asjad, M and Clerk, Aashish A and Massel, Francesco and Woolley, M J and Sillanp{\"a}{\"a}, Mika A},
  year = 2018,
  month = apr,
  journal = {Nature},
  volume = {556},
  number = {7702},
  pages = {478--482},
  doi = {10.1038/s41586-018-0038-x},
  langid = {english}
}

@article{onsager_reciprocal_1931,
  title = {Reciprocal {{Relations}} in {{Irreversible Processes}}. {{I}}.},
  author = {Onsager, Lars},
  year = 1931,
  month = feb,
  journal = {Physical Review},
  volume = {37},
  number = {4},
  pages = {405--426},
  publisher = {American Physical Society},
  doi = {10.1103/PhysRev.37.405},
  urldate = {2026-02-16}
}

@article{palomaki_entangling_2013,
  title = {Entangling Mechanical Motion with Microwave Fields.},
  author = {Palomaki, T A and Teufel, J D and Simmonds, R W and Lehnert, K W},
  year = 2013,
  month = nov,
  journal = {Science},
  volume = {342},
  number = {6159},
  pages = {710--713},
  doi = {10.1126/science.1244563},
  langid = {english}
}

@article{penrose_gravitys_1996,
  title = {On Gravity's Role in Quantum State Reduction},
  author = {Penrose, R},
  year = 1996,
  journal = {General relativity and gravitation},
  volume = {28},
  pages = {581}
}

@article{pirkkalainen_squeezing_2015,
  title = {Squeezing of {{Quantum Noise}} of {{Motion}} in a {{Micromechanical Resonator}}},
  author = {Pirkkalainen, J M and Damsk{\"a}gg, E and Brandt, M and Massel, Francesco and Sillanp{\"a}{\"a}, Mika A},
  year = 2015,
  month = dec,
  journal = {Physical Review Letters},
  volume = {115},
  number = {24},
  pages = {243601},
  doi = {10.1103/physrevlett.115.243601},
  langid = {english}
}

@article{pokharel_coupling_2022,
  title = {Coupling {{Capacitively Distinct Mechanical Resonators}} for {{Room-Temperature Phonon-Cavity Electromechanics}}},
  author = {Pokharel, Alok and Xu, Hao and Venkatachalam, Srisaran and Collin, Eddy and Zhou, Xin},
  year = 2022,
  month = sep,
  journal = {Nano Letters},
  volume = {22},
  number = {18},
  pages = {7351--7357},
  publisher = {American Chemical Society},
  issn = {1530-6984},
  doi = {10.1021/acs.nanolett.2c01848},
  urldate = {2024-08-29}
}

@article{poyatos_quantum_1996,
  title = {Quantum {{Reservoir Engineering}} with {{Laser Cooled Trapped Ions}}},
  author = {Poyatos, J F and Cirac, J I and Zoller, P},
  year = 1996,
  month = dec,
  journal = {Physical Review Letters},
  volume = {77},
  number = {23},
  pages = {4728--4731},
  doi = {10.1103/physrevlett.77.4728},
  langid = {english}
}

@article{tan_achieving_2013,
  title = {Achieving Steady-State Entanglement of Remote Micromechanical Oscillators by Cascaded Cavity Coupling},
  author = {Tan, Huatang and Buchmann, L. F. and Seok, H. and Li, Gaoxiang},
  year = 2013,
  month = feb,
  journal = {Physical Review A},
  volume = {87},
  number = {2},
  pages = {022318},
  issn = {1050-2947, 1094-1622},
  doi = {10.1103/PhysRevA.87.022318},
  urldate = {2026-04-20},
  copyright = {http://link.aps.org/licenses/aps-default-license},
  langid = {english}
}

@article{teufel_sideband_2011-1,
  title = {Sideband Cooling of Micromechanical Motion to the Quantum Ground State},
  author = {Teufel, J. D. and Donner, T. and Li, Dale and Harlow, J. W. and Allman, M. S. and Cicak, K. and Sirois, A. J. and Whittaker, J. D. and Lehnert, K. W. and Simmonds, R. W.},
  year = 2011,
  month = jul,
  journal = {Nature},
  volume = {475},
  number = {7356},
  pages = {359--363},
  publisher = {Nature Publishing Group},
  issn = {1476-4687},
  doi = {10.1038/nature10261},
  urldate = {2024-09-16},
  copyright = {2011 Springer Nature Limited},
  langid = {english}
}

@article{tsang_coherent_2010,
  title = {Coherent {{Quantum-Noise Cancellation}} for {{Optomechanical Sensors}}},
  author = {Tsang, Mankei and Caves, Carlton M.},
  year = 2010,
  month = sep,
  journal = {Physical Review Letters},
  volume = {105},
  number = {12},
  pages = {123601},
  publisher = {American Physical Society},
  doi = {10.1103/PhysRevLett.105.123601},
  urldate = {2026-03-11}
}

@article{tsang_evading_2012,
  title = {Evading {{Quantum Mechanics}}: {{Engineering}} a {{Classical Subsystem}} within a {{Quantum Environment}}},
  author = {Tsang, Mankei and Caves, Carlton M},
  year = 2012,
  month = sep,
  journal = {Physical Review X},
  volume = {2},
  number = {3},
  pages = {031016},
  doi = {10.1103/physrevx.2.031016},
  langid = {english}
}

@article{wang_reservoir-engineered_2013,
  title = {Reservoir-{{Engineered Entanglement}} in {{Optomechanical Systems}}},
  author = {Wang, Ying-Dan and Clerk, Aashish A.},
  year = 2013,
  journal = {Physical Review Letters},
  volume = {110},
  number = {25},
  pages = {253601},
  issn = {0031-9007},
  doi = {10.1103/physrevlett.110.253601},
  pmid = {23829736}
}

@article{wollman_quantum_2015,
  title = {Quantum Squeezing of Motion in a Mechanical Resonator},
  author = {Wollman, E E and Lei, C U and Weinstein, A J and Suh, J and Kronwald, A and Marquardt, F and Clerk, Aashish A and Schwab, K C},
  year = 2015,
  month = aug,
  journal = {Science},
  volume = {349},
  number = {6251},
  pages = {952--955},
  doi = {10.1126/science.aac5138},
  langid = {english}
}

@article{woolley_two-mode_2014,
  title = {Two-Mode Squeezed States in Cavity Optomechanics via Engineering of a Single Reservoir},
  author = {Woolley, M. J. and Clerk, A. A.},
  year = 2014,
  month = jun,
  journal = {Physical Review A},
  volume = {89},
  number = {6},
  pages = {063805},
  publisher = {American Physical Society},
  doi = {10.1103/PhysRevA.89.063805},
  urldate = {2026-04-24}
}

@article{xinexperiment,
  title = {In Preparation},
  author = {{Massel, Francesco} and {Zhou, Xin}},
  year = 2026,
  journal = {In preparation}
}

@article{zurek_decoherence_2003,
  title = {Decoherence, Einselection, and the Quantum Origins of the Classical},
  author = {Zurek, Wojciech Hubert},
  year = 2003,
  month = may,
  journal = {Reviews of Modern Physics},
  volume = {75},
  number = {3},
  pages = {715--775},
  doi = {10.1103/revmodphys.75.715},
  langid = {english}
}
\end{document}


\title{Quantum limits on squeezing - Supplemental Material}
\author{Francesco Massel}
\maketitle
In this Supplement we use $\Delta X^2$ for variances; thus $\Delta X^2\equiv \mathrm{Var}(X)$ and $\mathrm{Cov}(A,B)\equiv \frac12\langle AB+BA\rangle-\langle A\rangle\langle B\rangle$.
\section{Commutator budget and Lyapunov equation}
\label{sec:SI-1}

As a first step, we solve the LTI system describing the dynamics of the system
\begin{equation}
  \label{eq:1}
  \dot{\xi} = \bar{A} \xi + D \xi_\mathrm{in}
\end{equation}
as
\begin{equation}
  \label{eq:2}
  \xi(t)=e^{\bar A t}\xi(0)+\int_0^t e^{\bar A (t-\tau)}D\,\xi_\mathrm{in}(\tau)\,d\tau,
\end{equation}
which, for $t\to\infty$, can be written as ($\bar{A}$ stable)
\begin{equation}
  \label{eq:3}
  \xi(t)=\int_0^\infty e^{\bar A s}D\,\xi_{\rm in}(t-s)\,ds
  \quad (s=t-\tau).
\end{equation}
Assuming that the input fields satisfy (Markov property)
\begin{equation}
  \label{eq:4}
  [\xi(t),\xi^\dagger(t)]
= \int_0^\infty \int_0^\infty e^{\bar A s}D
[\xi_{\rm in}(t-s),\xi_{\rm in}^\dagger(t-s')]
D^\dagger e^{\bar A^\dagger s'}\,ds\,ds',
\end{equation}
and that the input fields are $\delta-$correlated,
\begin{equation}
  \label{eq:5}
  [xi_{\rm in}(t-s),\xi_{\rm in}^\dagger(t-s')]=\Sigma\delta(s-s')
\end{equation}
($\Sigma = [1,\, 0; 0,\, \text{-}1]$), we obtain
\begin{equation}
  \label{eq:6}
  [\xi(t),\xi^\dagger(t)]
    = \int_0^\infty e^{\bar A s}(D\Sigma D^\dagger)e^{\bar A^\dagger s}ds.
\end{equation}
If we pick the $i-$th input channel from Eq.~\eqref{eq:6} we obtain Eq. 1 of the main text
\begin{equation}
  \label{eq:7}
  W_i = \int_0^\infty e^{\bar{A}t} (D_{i}\Sigma D_{i}^\dag) e^{\bar{A}^\dag t} dt,
\end{equation}
where, as discussed in the main text, $W_i$ is a matrix whose elements $(W_i)_{jk}$ correspond to the portion of the commutator (``commutator budget'') $[\xi_k(t),\xi_k^\dagger(t)]$ that can be ascribed to the commutator of input mode $i$.
Defining $ F(t)=e^{\bar{A}t} (D_{i}\Sigma D_{i}^\dag) e^{\bar{A}^\dag t}$ and differentiating  Eq.~(\ref{eq:7}) with respect to time, we get
\begin{equation}
  \label{eq:8}
  \int_0^\infty \frac{dF}{dt},dt
  =\int_0^\infty \left(\bar A F(t)+F(t)\bar A^\dagger\right)dt
  = \bar A\int_0^\infty F(t)\,dt+\left(\int_0^\infty F(t)dt\right)\bar A^\dagger
=  \bar A W_i + W_i\bar A^\dagger.
\end{equation}
On the other hand, we have that
\begin{equation}
  \label{eq:9}
  \int_0^\infty \frac{dF}{dt}\,dt = F(\infty)-F(0) = - D_i\Sigma D_i^\dagger
\end{equation}
since $e^{\bar A t}\to 0$ and $e^{\bar A^\dagger t}\to 0$, allowing us to conclude that
\begin{equation}
  \label{eq:10}
  \bar A W_i + W_i \bar A^\dagger + D_i \Sigma D_i^\dagger=0.
\end{equation}

\section{Commutator budget sum rules.}
\label{sec:comm-budg-sum}
As discussed in the main text the relation $\sum_i (K_i)_{jk}=\delta_{jk}$ is straightforward. In this section we will prove, under the appropriate conditions the following relations.
\begin{enumerate}
  \item $\sum_j \gamma_j (K_i)_{jj}=\gamma_i$
  \item $\gamma_j (K_i)_{jj} = \gamma_i(K_j)_{ii}$
\end{enumerate}

\subsection{Weighted sum rule for the commutator channels}
\label{sec:sum_1}

As discussed in the text, in order to prove the relation $\sum_j \gamma_j (K_i)_{jj}=\gamma_i$ in addition to PR and stability, the following conditions are required:
\begin{enumerate}
  \item \textbf{No internal losses.} All damping is associated with an input (noise) port.
  \item \textbf{Diagonal dissipation/local input coupling.} the dissipative part of $\bar A$ is diagonal with rates \(\gamma_j\), and the port couplings are local:
  \begin{equation}
    \label{eq:11}
    D=\mathrm{diag}(\sqrt{\gamma_1},\dots,\sqrt{\gamma_N}) \quad\text{(in annihilation sector)}.
  \end{equation}
  \end{enumerate}

\paragraph*{Proof}
\label{sec:proof}
If we take the annihilation-sector Lyapunov equation
\begin{equation}
  \label{eq:12}
  A K_i + K_i A^\dagger + B_iB_i^\dagger=0,
\end{equation}
we can write its trace as
\begin{equation}
  \label{eq:13}
  \Tr[K_i (A + A^\dagger) + B_i B_i^\dagger ] =0
\end{equation}
which, considering diagonal-only dissipation  ($A+A^\dagger=-\mathrm{diag}(\gamma_j) $) and no hidden losses, can be written as
\begin{equation}
  \label{eq:14}
   \Tr[K_i \mathrm{diag}(\gamma_j)] =-\Tr[B_i B_i^\dagger]
\end{equation}
which can be written as
\begin{equation}
  \label{eq:15}
  \sum_j \gamma_j (K_i)_{jj} = \gamma_i
\end{equation}

\subsection{Reciprocity }
We show here that the reciprocity condition corresponds to the following relation $\gamma_j (K_i)_{jj} = \gamma_i(K_j)_{ii}$

From the reduced Lyapunov equation for $K_i$ we can deduce that $K_i$ can be written as
\begin{equation}
  \label{eq:16}
  K_i=\int_0^\infty e^{At}B_iB_i^\dagger\,e^{A^\dagger t}dt,
\end{equation}
and for local input coupling ($B_i=\sqrt{\gamma_i}\,e_i$)
\begin{equation}
  \label{eq:17}
   (K_i)_{jj}=\sum_{m,q}\int_0^\infty (e^{At})_{jm} \delta_{mi}\gamma_i \delta_{iq}\,(e^{A^\dagger t})_{qj} dt
\end{equation}
since $A^T=A$ for a reciprocal setup, we have
\begin{equation}
  \label{eq:18}
   (K_i)_{jj}=\gamma_i \int_0^\infty |(e^{At})_{ji}|^2 dt.
 \end{equation}
 Swapping the indexes $i \leftrightarrow j$, we have
 \begin{equation}
   \label{eq:19}
    (K_j)_{ii}=\gamma_j \int_0^\infty |(e^{At})_{ij}|^2 dt
 \end{equation}
 given, again, the relation $A^T=A$, we can write
 $(e^{At})_{ij}=(e^{At})_{ji}$ and therefore
 \begin{equation}
   \label{eq:20}
   \gamma_j(K_i)_{jj} =\gamma_i(K_j)_{ii}
 \end{equation}

\section{Cross bound for beam-splitter}
\label{sec:cross-bound-beam}

For the case of two-mode passive dynamics (with the same assumptions as above for dissipation and inputs), we can write
\begin{equation}
  \label{eq:21}
  \dot a = A a + B a_{\rm in},
\end{equation}
with
$$
A=\begin{pmatrix}-\gamma_1/2 & -iJ\\ -iJ & -\gamma_2/2\end{pmatrix},\quad
B=\mathrm{diag}(\sqrt{\gamma_1},\sqrt{\gamma_2}),
$$
and $J\in\mathbb R$, $\gamma_{1,2}>0$.
We can therefore write
\[
I_{ij} \doteq (K_j)_{ii}, \qquad K_j=\int_0^\infty e^{At}B_jB_j^\dagger e^{A^\dagger t}dt,\quad B_j=\sqrt{\gamma_j}e_j.
\]
In particular
 the cross weights  are \(I_{21}=(K_1)_{22}\) and \(I_{12}=(K_2)_{11}\), and $I_x$ is
$$
I_x \doteq \frac{I_{21}}{\gamma_1}=\frac{I_{12}}{\gamma_2}.
$$
With these definitions, the solution of the Lyapunov equation gives
\begin{equation}
  \label{eq:22}
  I_{21}=\frac{4 J^2 \gamma_1}{\left(\gamma_1+\gamma_2\right)\left(\gamma_1\gamma_2+4J^2\right)}
\end{equation}
and, similarly, $I_{12}=\frac{4J^2 \gamma_2 }{\left(\gamma_1+\gamma_2\right)\left(\gamma_1\gamma_2+4J^2\right)}$ and therefore
\begin{equation}
  \label{eq:23}
  I_x \doteq \frac{4 J^2}{\left(\gamma_1+\gamma_2\right)\left(\gamma_1\gamma_2+4J^2\right)}.
\end{equation}

\section{Variance bounds and Bogoliubov transformation}
\label{sec:vari-bounds-bogol}

In the derivation of Eq.~(4) of the main text, we exploit the fact that in the Bogoliubov frame the inequality
\begin{equation}
  \label{eq:24}
  \Delta X_1^2(\theta)/ \Delta X_{1,in}^2 + \Delta X_{\alpha}^2(\theta)/\Delta X_{\alpha,in}^2(\theta) \geq 1
\end{equation}
is valid for any $\theta$ and is minimized for some specific value $\theta_{min}$ ($X_1(\theta_{min})=X_{min,1}$, $X_\alpha(\theta_{min})=X_{min,\alpha}$).
The application of the (inverse) Bogoliubov transformation
$ a_2= \alpha \cosh[\xi] - \alpha^\dagger \sinh[\xi] $ to quadrature operators gives
$$
 X_{min,2,(in)}= X_{min,\alpha,(in)}\exp [-\xi]
$$
and therefore
\begin{equation}
  \label{eq:25}
  (\Delta X_{2,(in)}^2)_{min} = (\Delta X_{\alpha,(in)}^2)_{min}\exp[-2\xi] \qquad  \text{and, trivially} \,(\Delta X_{1,(in)}^2)_{min} =(\Delta X_{1,(in)}^2)_{min}.
\end{equation}
 For the 3-mode case, a similar analysis can be performed to obtain Eq.~(12) of the main text.

\section{Exact variance Lyapunov equations for dissipative squeezing. }
\label{sec:exact-lyap-equat}

In this section, we solve the Lyapunov equations associated with the quadrature EOMs given in Eqs. (6,7) of the main text. Defining input correlators $\langle
D_1 =X_{1,\mathrm{in}}(t)X_{1,\mathrm{in}}(t') \rangle = \langle
Y_{1,\mathrm{in}}(t)Y_{1,\mathrm{in}}(t') \rangle=
N_\mathrm{1}$, where $N_\mathrm{1} =
\left(n_\mathrm{1,th}+ \frac{1}{2}\right)$ and $n_\mathrm{1,th}$ is the
thermal occupation for mode $a_1$. Similar equations hold for mode $a_2$.
Under these hypotheses, the steady-state covariance matrix $V_\mathrm{x}$ is a
solution of the continuous-time Lyapunov equation

\begin{align}
  \label{eq:26}
  A_\mathrm{x} V_\mathrm{x} +  V_\mathrm{x} A_\mathrm{x}^T + D_\mathrm{x} =0 \, \qquad \mathrm{x}=1,2
\end{align}

The solution for block 1 gives
\begin{align}
  \label{eq:27}
  \mathrm{Cov}(X_1,Y_2)
   &=  2 \frac{G_\Delta \gamma_2 N_2 \gamma_{1-} - G_\Sigma \gamma_1 N_1 \gamma_{2+}}
  {(\gamma_{1-}+\gamma_{2+}) \left(\gamma_{1-} \gamma_{2+}+ 4 G_\Delta G_\Sigma\right)} \\
  \label{eq:28}
  \Delta X_1^2 &= \frac{\gamma_1}{\gamma_{1-}} N_1 + \frac{ 2 G_\Delta}{\gamma_{1-}}  \mathrm{Cov}(X_1,Y_2)   \\
   \label{eq:29}
  \Delta Y_2^2 &=  \frac{\gamma_2}{\gamma_{2+}} N_2 - \frac{ 2 G_\Sigma}{\gamma_{2+}}  \mathrm{Cov}(X_1,Y_2)
\end{align}
while for block 2 we have
\begin{align}
  \label{eq:30}
   \mathrm{Cov}(X_2,Y_1)
   & = 2 \frac{G_\Delta \gamma_1 N_1 \gamma_{2-} - G_\Sigma \gamma_2 N_2 \gamma_{1+}}  {(\gamma_{1+}+\gamma_{2-}) \left(\gamma_{1+} \gamma_{2-}+ 4 G_\Delta G_\Sigma\right)} \\
  \label{eq:31}
  \Delta X_2^2&= \frac{\gamma_2}{\gamma_{2-}} N_2 + \frac{ 2 G_\Delta}{\gamma_{2-}}  \mathrm{Cov}(X_2,Y_1)   \\
   \label{eq:32}
  \Delta Y_1^2&=  \frac{\gamma_1}{\gamma_{1+}} N_1 - \frac{ 2 G_\Sigma}{\gamma_{1+}}  \mathrm{Cov}(X_2,Y_1).
\end{align}

\paragraph*{Dissipative squeezing.} For purely dissipative squeezing (i.e.
$\eta_\mathrm{j}=0$, it is possible to recast the expression for the quadrature
fluctuations as
\begin{align}
  \label{eq:33}
  \mathrm{Cov}(X_1,Y_2)
   &= \frac{2 \gamma_1 \gamma_{2}  \left(G_\Delta N_2 - G_\Sigma N_1\right)}
  {(\gamma_{1}+\gamma_{2}) \left(\gamma_{1} \gamma_{2}+ 4 G_\Delta G_\Sigma\right)}\\
  \label{eq:34}
  \Delta X_1^2 &= N_1 + \frac{ 2 G_\Delta}{\gamma_{1}}  \mathrm{Cov}(X_1,Y_2)   \\
   \label{eq:35}
  \Delta Y_2^2 &=  N_2 - \frac{ 2 G_\Sigma}{\gamma_{2}}  \mathrm{Cov}(X_1,Y_2),
\end{align}
where analogous expressions hold for block 2 ($X_2$,$Y_1$).
With the substitutions $G_\Sigma \to \mathcal{G} \exp\left[ \chi \right]$,
$G_\Delta \to \mathcal{G} \exp\left[ -\chi \right]$, we can see that the two squeezed
quadratures for modes $1$ and $2$ are
\begin{align}
  \label{eq:36}
    \Delta X_1^2 &=  \frac{\gamma_1 N_1 \left[\gamma_2 \left(\gamma_1+ \gamma_2\right) + 4 \mathcal{G}^2\right]+ 4 \gamma_2 \mathcal{G}^2 e^{-2 \chi} N_2}
  {(\gamma_{1}+\gamma_{2}) \left(\gamma_{1} \gamma_{2}+ 4 \mathcal{G}^2\right)}   \\
   \label{eq:37}
     \Delta X_2^2 &=  \frac{\gamma_2 N_2 \left[\gamma_1 \left(\gamma_1+ \gamma_2\right) + 4 \mathcal{G}^2\right]+ 4 \gamma_1 \mathcal{G}^2 e^{-2 \chi} N_1}
  {(\gamma_{1}+\gamma_{2}) \left(\gamma_{1} \gamma_{2}+ 4  \mathcal{G}^2 \right)}.
\end{align}
These variances are used plot the squeezing power in Fig.~1 of the main text.
For completeness, the two anti-squeezed quadratures are given by
\begin{align}
  \label{eq:38}
    \Delta Y_1^2 &=  \frac{\gamma_1 N_1 \left[\gamma_2 \left(\gamma_1+ \gamma_2\right) + 4 \mathcal{G}^2\right]+ 4 \gamma_2 \mathcal{G}^2 e^{2 \chi} N_2}
  {(\gamma_{1}+\gamma_{2}) \left(\gamma_{1} \gamma_{2}+ 4  \mathcal{G}^2\right)}  \\
   \label{eq:39}
     \Delta Y_2^2 &=  \frac{\gamma_2 N_2 \left[\gamma_1 \left(\gamma_1+ \gamma_2\right) + 4 \mathcal{G}^2\right]+ 4 \gamma_1 \mathcal{G}^2 e^{2 \chi} N_1}
  {(\gamma_{1}+\gamma_{2}) \left(\gamma_{1} \gamma_{2}+ 4  \mathcal{G}^2 \right)}.
\end{align}
The smallest possible value for $ \Delta X_1^2$ and
$\Delta X_2^2$ is reached in the limit $\mathcal{G}, \xi \to \infty $,
in that case we have that
\begin{subequations}
  \begin{align}
    \label{eq:40}
    \Delta X_1^2 = \frac{\gamma_1}{\gamma_1 + \gamma_2} N_1\\
    \label{eq:41}
     \Delta X_2^2 = \frac{\gamma_2}{\gamma_1 + \gamma_2} N_2.
  \end{align}
\end{subequations}
Eqs. (\ref{eq:40}-\ref{eq:41}) signal two main conclusions:

\begin{enumerate}
  \item as long as $\gamma_1 \gg \gamma_2$, it is possible to reduce the
        fluctuations for the best-squeezed quadratures of one mode (mode $2$
        in this case) to arbitrarily small values. This case corresponds to the conditions usually encountered in optomechanics;
  \item the overall ``squeezing power''
        $\Delta X_1^2/\Delta X_{i,\mathrm{in}}^2 + \Delta X_2^2/ \Delta X_{i,\mathrm{in}}^2$
        cannot be reduced below 1.
\end{enumerate}

\paragraph*{Dissipative + parametric squeezing.}

In the presence of both dissipative and parametric squeezing, the general
expression for the quadrature variances are given by
\begin{subequations}
  \begin{align}
    \label{eq:42}
   \Delta X_1^2= \frac{\gamma_1}{\gamma_{1\,-}} N_1 +
                           \frac{4 G_\Delta}{\gamma_{1\,-}}
                            \frac{\gamma_{1\,-} G_\Delta\gamma_2 N_2 - \gamma_{2\,+} G_\Sigma \gamma_1 N_1}
    {\left(\gamma_{1\,-}+\gamma_{2\,+}\right)\left(\gamma_{1\,-}\gamma_{2\,+}+4 G_\Delta G_\Sigma\right)} \\
    \Delta X_2^2= \frac{\gamma_2}{\gamma_{2\,-}} N_2 +
                           \frac{4 G_\Delta}{\gamma_{2\,-}}
                            \frac{\gamma_{2\,-} G_\Delta\gamma_1 N_1 - \gamma_{1\,+} G_\Sigma \gamma_2 N_2}
    {\left(\gamma_{1\,+}+\gamma_{2\,-}\right)\left(\gamma_{1\,+}\gamma_{2\,-}+4 G_\Delta G_\Sigma\right)} \\
   \Delta Y_1^2= \frac{\gamma_1}{\gamma_{1\,+}} N_1 +
                            \frac{4 G_\Sigma}{\gamma_{1\,+}}
                            \frac{\gamma_{1\,+} G_\Sigma\gamma_2 N_2 - \gamma_{2\,-} G_\Delta \gamma_1 N_1}
    {\left(\gamma_{1\,+}+\gamma_{2\,-}\right)\left(\gamma_{1\,+}\gamma_{2\,-}+4 G_\Sigma G_\Delta\right)} \\
     \Delta Y_2^2 = \frac{\gamma_2}{\gamma_{2\,+}} N_2 +
                            \frac{4 G_\Sigma}{\gamma_{2\,+}}
                            \frac{\gamma_{2\,+} G_\Sigma\gamma_1 N_1 - \gamma_{1\,-} G_\Delta \gamma_2 N_2}
    {\left(\gamma_{1\,-}+\gamma_{2\,+}\right)\left(\gamma_{1\,-}\gamma_{2\,+}+4 G_\Sigma G_\Delta\right)}
  \end{align}
\end{subequations}
 If we allow for $\eta_\mathrm{i} \neq 0 $, the limits described by
Eqs.(\ref{eq:31}-\ref{eq:32}) are altered. In particular the
 the expressions for $\Delta X_1^2$ and $ \Delta X_2^2$ in the limit $\mathcal{G},
\xi \to \infty $, become
\begin{subequations}
  \begin{align}
    \label{eq:43}
   \Delta X_1^2= \frac{\gamma_1}{\gamma_{1\,-} + \gamma_{2\,+}} N_1\\
    \label{eq:44}
   \Delta X_2^2= \frac{\gamma_2}{\gamma_{2\,-} + \gamma_{1\,+}} N_2.
  \end{align}
\end{subequations}

Remembering that $\gamma_{1,2\,\pm}=\gamma_{1,2} \pm \eta_{1,2}$, we have that
$\Delta X_1^2<1/2$ requires us to have $\eta_{2}-\eta_{1} > \gamma_1 - \gamma_2$,
while $\Delta X_2^2<1/2$ requires $\eta_{2}-\eta_{1} < \gamma_1 -
\gamma_2$. We can therefore conclude that for $\eta_{1,2} \in \mathbb{R}$,
either positive or negative, one quadrature fluctuation is required to be above
the -3 dB limit. Interestingly, the overall squeezing power for $G_\pm,
\eta_\mathrm{j} \neq 0$
\begin{align}
  \label{eq:45}
   \frac{\Delta X_1^2}{\Delta X_1^2|G\pm =0} + \frac{\Delta X_2^2}{\Delta X_2^2|G\pm =0} = \frac{\left(\gamma_1 + \gamma_2 \right)^2 - (\eta_1-\eta_2)(\gamma_1-\gamma_2)}{\left(\gamma_1 + \gamma_2 \right)^2- (\eta_1-\eta_2)^2}
\end{align}
can be improved with respect to the purely dissipative case, reaching the optimal
value $1/2 + \sqrt{\gamma_1 \gamma_2}/(\gamma_1 + \gamma_2)$ for $\eta_1 -\eta_2
= (\gamma_1^2-\gamma_2^2)/(\sqrt{\gamma_2}+\sqrt{\gamma_1})^2$. This is the situation depicted in Fig.~(2) of the main text.
In the limit of either $\gamma_{1,2} \to 0$ the overall squeezing power reaches a value $1/2^+$ (i.e. is bounded by $1/2$ from below) . For this case, one quadrature  can be squeezed to zero, while for the other the fluctuations are reduced by a factor $1/2$.